\documentclass[iop]{emulateapj}
\usepackage{graphicx}
\usepackage{epsfig}
\usepackage{epstopdf}

\begin{document}


\shorttitle{Torsional motions in spicules}
\shortauthors{De Pontieu et al.}
\title{Ubiquitous torsional motions in type~{\sc II} spicules}

\author{B. De Pontieu\altaffilmark{1}}
\email{bdp@lmsal.com}
\author{M. Carlsson\altaffilmark{2}}
\author{L.H.M. Rouppe van der Voort\altaffilmark{2}}   
\author{R.J. Rutten\altaffilmark{2}}
\author{V.H. Hansteen\altaffilmark{1,2}}
\author{H. Watanabe\altaffilmark{3}}


\affil{\altaffilmark{1}Lockheed Martin Solar \& Astrophysics Lab, Org.\ A021S,
  Bldg.\ 252, 3251 Hanover Street Palo Alto, CA~94304 USA}

\affil{\altaffilmark{2}Institute of Theoretical Astrophysics,
  University of Oslo,   %
  P.O. Box 1029 Blindern, N-0315 Oslo, Norway}

\affil{\altaffilmark{3}Unit of Synergetic Studies for Space, Kyoto University, 17 Kitakazan Ohmine-cho, Yamashina, Kyoto 607-8471}

\begin{abstract}
  Spicules are long, thin, highly dynamic features that jut out
  ubiquitously from the solar limb.  They dominate the interface
  between the chromosphere and corona and may provide significant
  mass and energy to the corona.  We use high-quality observations
  with the Swedish 1-m Solar Telescope to establish that so-called
  type {\sc II} spicules are characterized by the simultaneous action
  of three different types of motion: {\em (i)\/} field-aligned flows
  of order 50-100 km\,s$^{-1}$, {\em (ii)\/} swaying motions of order
  15-20 km\,s$^{-1}$, and {\em (iii)\/}, torsional motions of order
  25-30 km\,s$^{-1}$. The first two modes have been studied in detail
  before, but not the torsional motions.  Our analysis of many
  near-limb and off-limb spectra and narrow-band images using multiple
  spectral lines yields strong evidence that most, if not all, type-II
  spicules undergo large torsional modulation and that these motions,
  like spicule swaying, represent Alfv\'enic waves propagating outward
  at several hundred km\,s$^{-1}$.  The combined action of the
  different motions explains the similar morphology of spicule bushes 
  in the outer red and blue wings of chromospheric lines, and needs to be taken into
  account when interpreting Doppler
  motions to derive estimates for field-aligned flows in spicules and
  determining the Alfv\'enic wave energy in the solar atmosphere.  
  Our results also suggest that large torsional motion is an ingredient in
  the production of type-II spicules and that spicules play an
  important role in the transport of helicity
  through the solar atmosphere.
\end{abstract}

\section{Introduction} 
\label{sec:intro}


Between the photosphere and corona lies the chromosphere, a
region of relatively cool plasma that is most conspicuous
in the hydrogen H$\alpha$ Balmer line. The upper chromosphere is
dominated by spicules, thin jets of chromospheric plasma that reach
heights of $10\,000$~km or more above the photosphere. Although
spicules were described already by Secchi in 1877, 
understanding their physical nature has progressed only slowly
\citep[reviews
by][]{1968SoPh....3..367B,2000SoPh..196...79S}.
The launch of the Hinode satellite \citep{2007SoPh..243....3K}   
  %
and the combined use of adaptive optics and image post-processing 
\citep{2005SoPh..228..191V}   
in ground-based observing have revolutionized our view of spicules. 

There are (at least) two types \citep{2007PASJ...59S.655D}.
Type~{\sc i} spicules reach heights of 2-9~Mm, show up- and downward
velocities of 10--30~km\,s$^{-1}$, and have lifetimes of
3--10~minutes.
They probably correspond to on-disk dynamic fibrils caused by shock
waves that propagate upward into the chromosphere
\citep{2006ApJ...647L..73H,2007ApJ...660L.169R,
  2007ApJ...655..624D} \citep[see also][]{1995ApJ...450..411S}. Type~{\sc ii} spicules reach larger heights at
velocities of order 50--100~km\,s$^{-1}$.  They have shorter
lifetimes, of order 100\,s, and usually only exhibit upward motion
before their rapid fading in the chromospheric lines in which they are
detected.  
On the disk they appear as rapidly moving absorption features 
in the blue wings of chromospheric lines
\citep{2008ApJ...679L.167L, 2009ApJ...705..272R}. 
In this paper we focus on this intriguing class of features.

Our previous studies suggest that type~{\sc ii} spicules represent
impulsively accelerated chromospheric material that is continuously
heated while it rises
\citep{2009ApJ...701L...1D,2011Sci...331...55D}. The cause of the
heating and acceleration are unknown, but a magnetic process such as
reconnection and/or flux emergence is most likely
\citep{2007PASJ...59S.655D, 2010ApJ...722.1644S,   
  2011ApJ...736....9M}.   

Type~{\sc ii} spicules show other motions in addition to radial
outflow.  In the Ca~{\sc ii}~H line they are seen to sway to-and-from
transversely with amplitudes of order 10--20~km\,s$^{-1}$ and
periodicities of 100--500\,s, suggesting Alfv\'enic waves
\citep{2007Sci...318.1574D}.  The continuation of these motions in
transition region and coronal lines suggests that they may help drive the solar wind
\citep{2011Natur.475..477M}.  


Other types of motion are less well established.
\citet{2008ASPC..397...27S}   
reported that some spicules appear as double threads with evidence of
spinning motion.  \citet{2011A&A...532L...9C} and
\citet{2012arXiv1201.3199C}   
suggest that similar spinning explains the tilts of ultraviolet lines
in so-called explosive events producing larger-scale macro spicules.
Spectral-line tilts were noted earlier in observations at the
limb and also attributed to spicule rotation 
\citep{1972ARA&A..10...73B},   
or not interpreted \citep{1979RNAAS..82..223H}.   

In this paper we definitely confirm the indications for twisting
spicular motions by unequivocally detecting torsional spicule
modulation in limb spectroscopy and imaging spectroscopy with
unprecedented spectral and spatial resolution.  


\begin{figure*}[!t]
\includegraphics[width=0.5\textwidth]{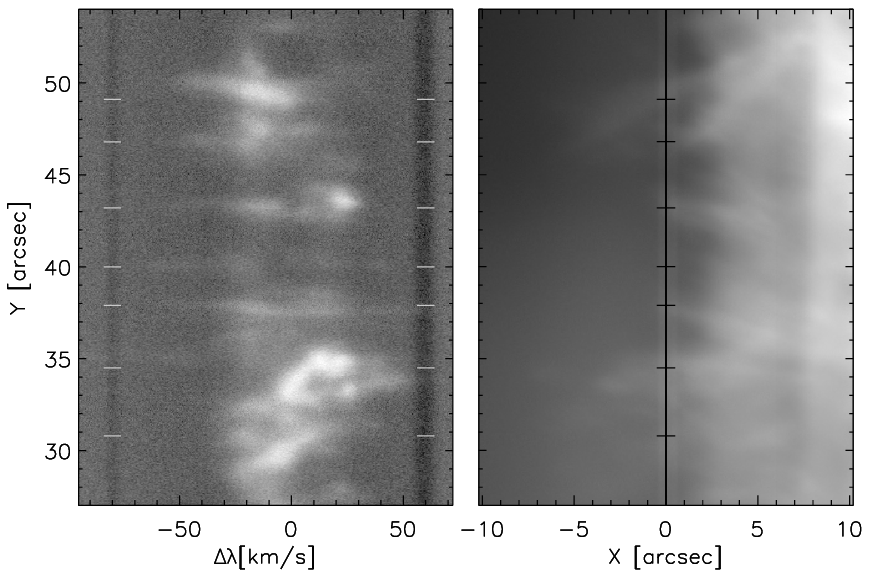}
\includegraphics[width=0.5\textwidth]{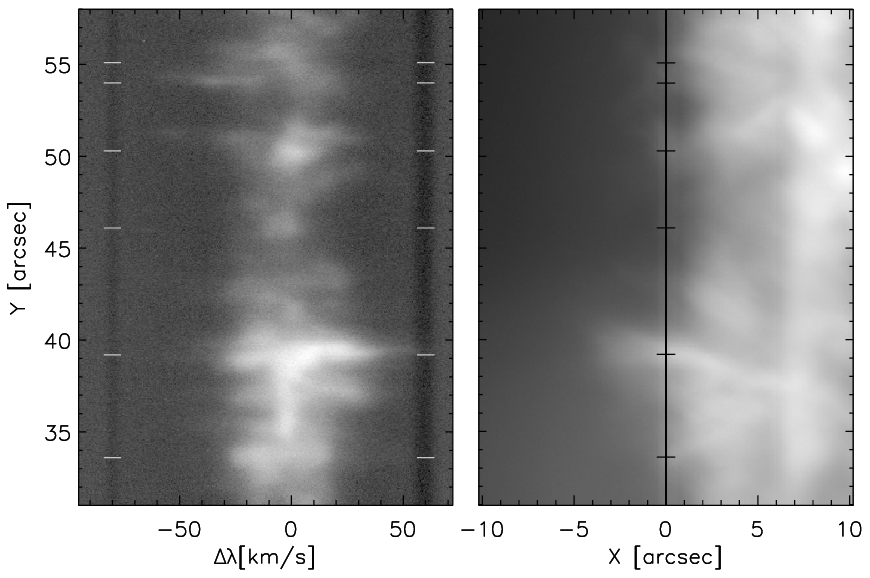}\par
\includegraphics[width=\textwidth]{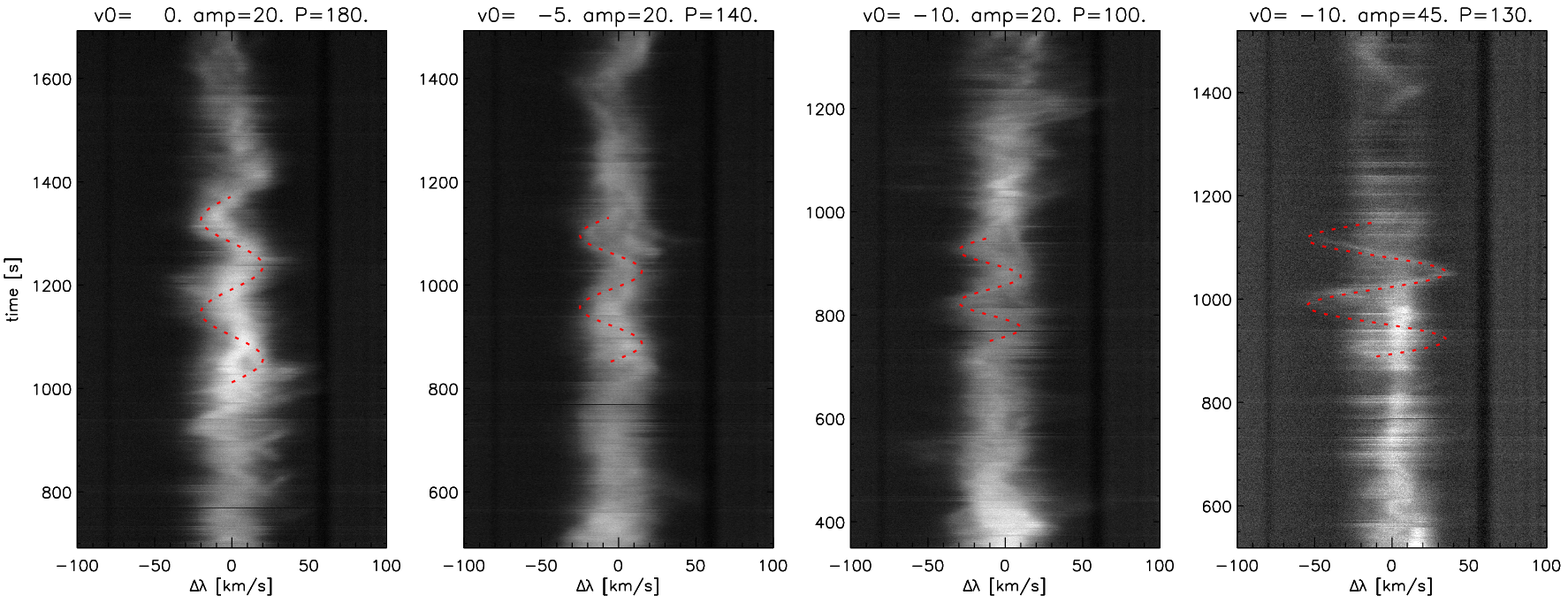}
\caption{Examples of large torsional motion in off-limb spicules
  in Ca\,{\sc II}\,H.  Top row shows the spatial variation of the
  Ca\,{\sc II}\,H profile, with the righthand panel of each pair a slit-jaw
  image with the off-limb intensities enhanced with a radial filter.
  These show spicules protruding upwards from the limb which is to the
  right. The lefthand panels are corresponding Ca\,{\sc II}\,H
  spectrograms.  The black line in the slit-jaw images shows the slit
  location.  The spectrograms are dominated by linear features that
  are often tilted from the horizontal dispersion direction, a clear
  signature of torsional motions, typically over 20-40~km\,s$^{-1}$.
  Selected spicules are marked by black ticks in the slit-jaw images
  and white ticks in the spectrograms. 
  Bottom row shows examples of large swaying motion in
  off-limb spicules in Ca\,{\sc II}\,H.  Each panel
  shows the temporal evolution of the Ca\,{\sc II}\,H profile.
  Despite
  significant superposition of multiple oscillatory signals there are
  often episodes of quasi-sinusoidal oscillation.  For guidance we have
  overplotted sinus curves with the amplitude (``amp'' in
  km\,s$^{-1}$) and period (P in s) specified in each legend.
\label{fig:lx}}
\end{figure*}

\section{Observations}
\label{sec:observations}

We analyze various data sets obtained with the Swedish 1-m Solar
Telescope \citep[SST,][]{2003SPIE.4853..341S}   
on La Palma: 
slit spectroscopy with the 
TRI-Port Polarimetric Echelle-Littrow spectrograph
\citep[TRIPPEL,][]{2011A&A...535A..14K}   
and imaging spectroscopy with the CRisp Imaging SpectroPolarimeter
\citep[CRISP,][]{2008ApJ...689L..69S}.  
All observations used adaptive optics.

With the TRIPPEL spectrograph we observed the Ca\,{\sc II}\,H line at 396.8~nm
with nominal spectral resolution of 1.6~pm.  A slit-jaw camera of the
same type was slaved to the spectrum camera. The data were binned
to 0.49~pm spectral and 0.068\arcsec\ spatial pixel sizes.
The spectrograms were corrected for dark current, gain variations, and
spectrograph distortions following \citet{2007ApJ...655..615L}.  We
use a 55\,min
duration, 0.91\,s cadence sequence of Ca\,{\sc II}\,H spectrograms taken on
2009 October 8 
in excellent seeing.  The slit was set parallel to the limb at various
heights above the limb ranging from 3\arcsec\ to 10\arcsec.
The exposure time was 800~ms.  

CRISP contains a dual Fabry-P{\'e}rot interferometer and allows for
fast ($<$50~ms) wavelength tuning within a spectral region.
High spatial resolution and precise alignment between the sequentially
taken images for different tunings is achieved with the
image restoration technique of 
\citet{2005SoPh..228..191V}   
and \citet{2009ApJ...705..272R}.   
In section~\ref{sec:results} we present results from two CRISP data
sets: a 17\,min 
sequence at the ultra-high cadence of 0.44\,s sampling
H\ensuremath{\alpha} (656.3~nm)
only at $\pm 1204$~m\AA\ ($\pm
55$~km\,s$^{-1}$), registering the limb near AR11230 
on 2011 June 11,
  %
and an earlier 36\,min, 
17\,s cadence sequence of finely spaced H\ensuremath{\alpha} and
Ca~{\sc ii} 854.2~nm profile samplings registering the limb
on 2010 June 27.   
\begin{figure}[!t]
\includegraphics[width=\columnwidth]{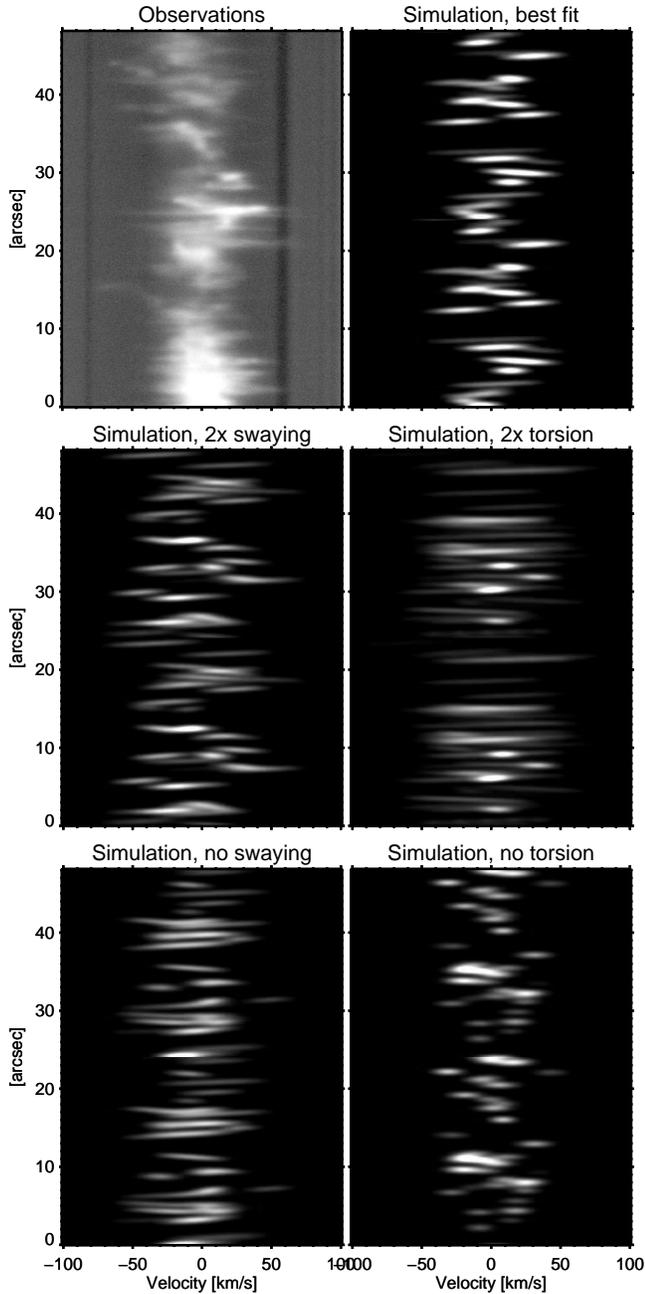}
\caption{Comparison between observations and Monte Carlo simulations
  in $\lambda\!-\!y$ displays.  The
  best-fit solution has field-aligned velocities of 
  60~km\,s$^{-1}$, torsional velocities of 30~km\,s$^{-1}$ and swaying
  motions of 15~km\,s$^{-1}$ with periods of 100-300\,s.
  Significant discrepancies occur for larger swaying, more torsion, no
  swaying, and no torsion.}
\label{fig:mc1}
\end{figure}

\begin{figure}[!t]
\includegraphics[width=\columnwidth]{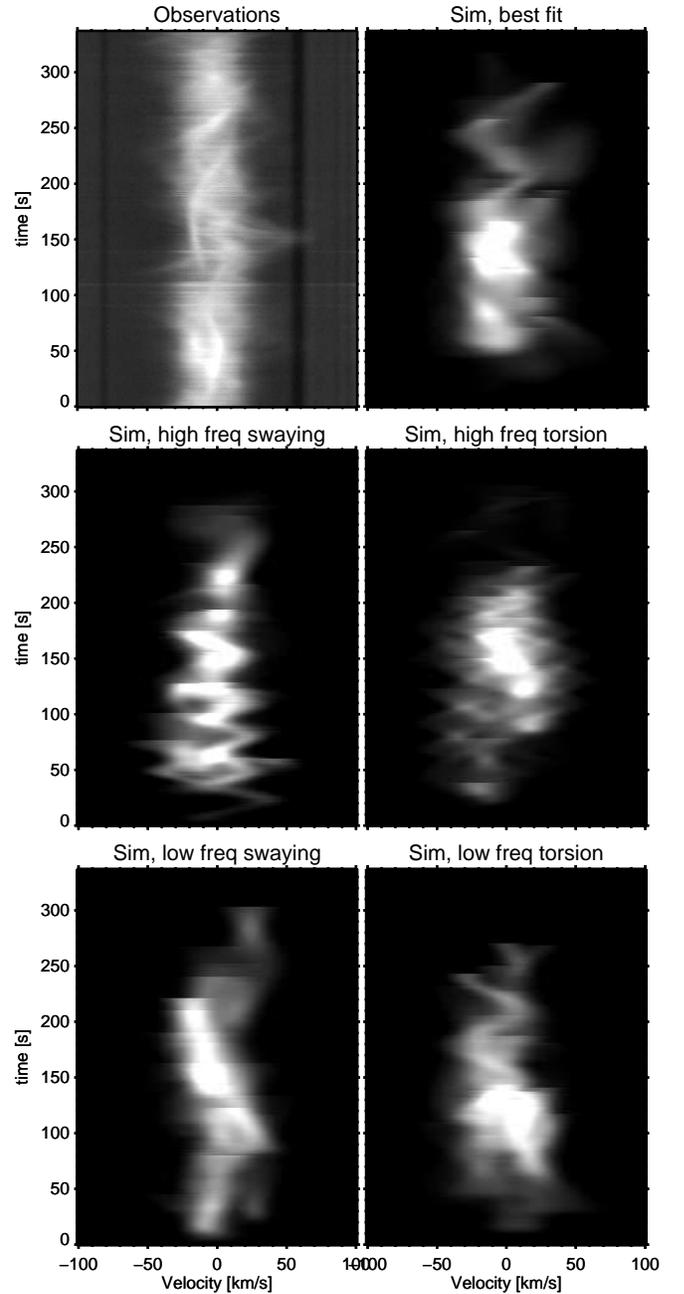}
\caption{Comparison between observations and Monte Carlo simulations
  in $\lambda\!-\!t$
  displays.  The
  best-fit solution has field-aligned velocities of 
  60~km\,s$^{-1}$, torsional velocities of 30~km\,s$^{-1}$ and swaying
  motions of 15~km\,s$^{-1}$ with periods of 100-300\,s.
  Significant discrepancies occur for wave periods that are much higher or lower
  than these values. }
\label{fig:mc2}
\end{figure}

\begin{figure*}[!t]
\includegraphics[bb=0 0 482 153, width=\textwidth]{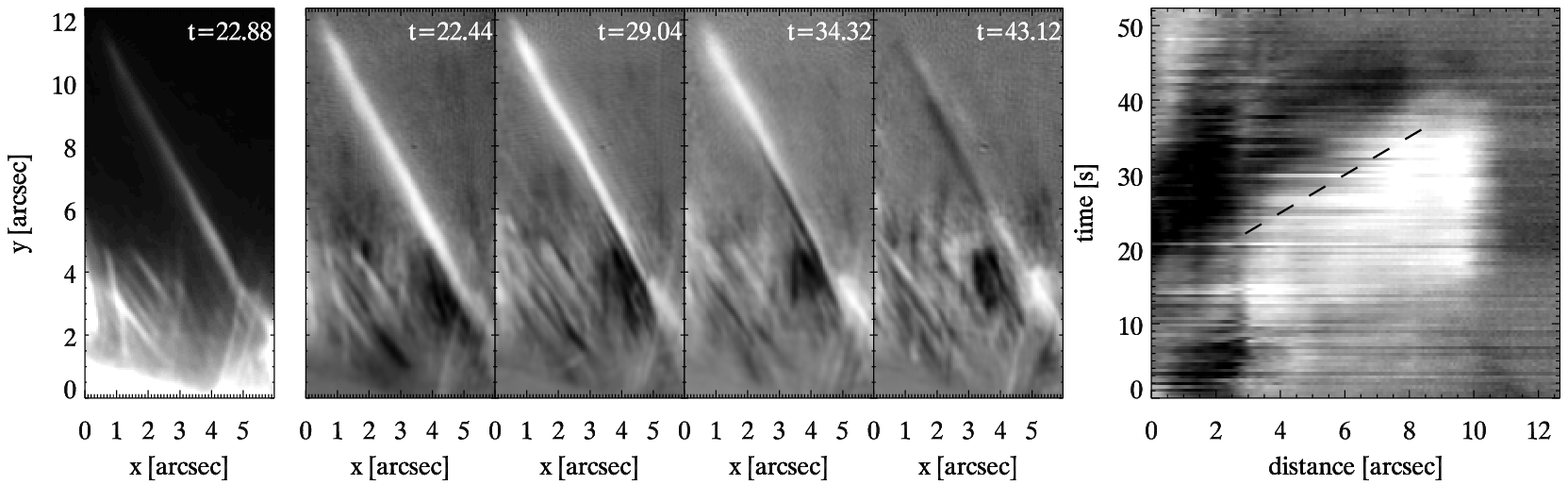}
\caption{Example of sign change in spicule Doppler tilt in the
  ultra-high cadence (0.44\,s) $\pm 55$~km\,s$^{-1}$ CRISP H$\alpha$
  data.  Lefthand panel shows an H$\alpha$ wing image at maximum
  spicule length. Center panels contain sequences of four Doppler
  images at different sampling times (in seconds, blueshift
  bright). The righthand panel show space-time diagrams measured along
  the spicule axis. Dashed guidelines illustrate a propagation speed of 285~km\,s$^{-1}$.}
\label{fig:dop}
\end{figure*}

\section{Results} 
\label{sec:results}

Our various observations all show clear evidence of torsional motions
on very small scales, of order 0.7\arcsec\ or less. We first
demonstrate these in the TRIPPEL spectra.  Since the slit was oriented
approximately parallel to the limb, it crossed many spicules. In
Figure~\ref{fig:lx} (top row) we show two $\lambda\!-\!y$ cuts, 
selecting moments when the slit was located at least several
arcseconds above the limb to avoid the enormous line-of-sight
superposition at and just above the limb.  The slit-jaw images indeed
show many distinct spicules that seem at least partially resolved.
Each spectral $\lambda\!-\!y$ panel contains multiple cases in which a
spicule appears as a thin bright streak across the spectrum, with a
small tilt from the horizontal wavelength direction and a substantial
offset from nominal line center.  We detected such behavior throughout
the 55-min sequence at all sampled heights.  The tilted-streak
morphology indicates relative redshift on one side of a spicule,
blueshift on the other.  This reversal in transverse motion is the
signature of torsional spicule motion.  In addition, the substantial
offsets of the tilted streaks from nominal line center can be
understood as the superposition of swaying motion and a projection of
field-aligned flow onto the line-of-sight.

Next, we examine how our data can be used to separate and constrain
the magnitudes and other properties of these three types of spicular
motion. Figure~\ref{fig:lx} (bottom row) shows the temporal variation of the Ca\,{\sc II}\,H profile.
Although significant line-of-sight superposition occurs, we find clear
evidence of quasi-sinusoidal episodes in the Dopplershifts.  To get an
idea of the typical amplitudes and periods of such oscillations we
overplot simple functions $v=v_0 +a \sin (2 \pi (t \!-\! t_0)/P)$ and
find that amplitudes of order 20-40 km\,s$^{-1}$ and periods of order
100-200\,s are common.  Swaying and torsional motions can both produce
such sinusoidal signature in $\lambda\!-\!t$ plots, but they have a
different appearance in a time sequence of successive $\lambda\!-\!y$
plots, with swaying motions showing wiggling of the entire streak-like
feature whereas torsional modulation shows up as a tilted streak that
shrinks into a more vertical feature and then becomes tilted with
opposite sign.  Detailed inspection of many such plots showed that
temporal undulation as in Figure~\ref{fig:lx} (bottom row) is sometimes associated
with swaying and sometimes with torsional modulation.

Our data inspection thus reveals evidence for three distinct
types of spicular motion.  Given the limitations of our data sets (in
particular the enormous line-of-sight superposition at the limb and
variations in seeing quality), we find that the best method to derive
statistical properties of these different motions is to compare the
observations with Monte Carlo simulations. We adapt methodology
used previously by 
\citet{2007Sci...318.1574D} 
and \citet{2008ApJ...673L.219M}: 
we consider $N$ spicules and impose on each a field-aligned flow
$v_p$, a torsional motion $v_t$, and a swaying motion $v_s$.
Inspired by our earlier analyses of type~{\sc ii} spicules we assume that, 
during their lifetime $T$, they continuously grow with constant
velocity $v_p$ along a straight path, inclined by an angle $\alpha$ from
the vertical and with an azimuth angle $\beta$, until they fade rapidly
from view.  The torsional and swaying motions are assumed to have
periods $P_t$ and $P_s$, respectively, with random phase.

Figures~\ref{fig:mc1} and \ref{fig:mc2} shows examples of comparisons between observed
$\lambda\!-\!y$ and $\lambda\!-\!t$ plots and results from these Monte
Carlo simulations for different parameter choices.  We made many such
plots with large parameter variation and defined a best-fit solution 
by selecting parameter combinations that reproduce the
observed behavior the best.  
Per spicule this best-fit choice takes $v_p$ randomly from a Gaussian
distribution around 60 km\,s$^{-1}$ with standard deviation $\sigma \!=\!
10$~km\,s$^{-1}$ (similar to what is found by Pereira et al., in
prep.), $T$ from a Gaussian distribution around 120\,s with
$\sigma\!=\!30$\,s (Pereira et al., in prep.), $\alpha$ from a Gaussian
distribution around 20 degrees with $\sigma\!=\!10$ degrees (based on
slit-jaw images), $\beta$ from a uniform distribution over
0-360~degrees, $v_t$ from a Gaussian distribution around
30~km\,s$^{-1}$ with $\sigma\!=\!10$~km\,s$^{-1}$, $v_s$ from a Gaussian
distribution around 15~km\,s$^{-1}$ with $\sigma\!=\!5$~km\,s$^{-1}$, and
both $P_s$ and $P_t$ from a uniform distribution over 100-300\,s.
Comparison of the $\lambda\!-\!y$ and $\lambda\!-\!t$ panels for this
best-fit solution with the observations in the top left panels
of Figures~\ref{fig:mc1} and \ref{fig:mc2} and in 
Figure~\ref{fig:lx} shows that the best-fit solution
reproduces, statistically, the appearance of both types of data, with
a similar multitude of slightly tilted streak-like features
in $\lambda\!-\!y$ and similar sinusoidal swings in $\lambda\!-\!t$.

How well defined are the best-fit parameters, given the large number
of free parameters in our simulations?  To answer this question we
ran multiple simulations in which we changed the distribution for
only one parameter, keeping all others fixed.  Examples are given in
the lower rows of Figures~\ref{fig:mc1} and \ref{fig:mc2}.  In the first panel of the second
row (Fig.~\ref{fig:mc1}) the mean swaying amplitude is doubled and gives an overall zig-zag
pattern with too many extremes.  Similarly, doubling the mean
torsional amplitude gives streaks that are too wide in velocity.
Removing the swaying (bottom-left panel) produces too little zig-zag
motion in $\lambda\!-\!y$.  No torsional motion gives streaks that are
too narrow in velocity.  Similarly, in the $\lambda\!-\!t$ 
panels of Fig.~\ref{fig:mc2}, doubling the frequencies to periods of 50-100\,s of
the swaying or torsional motion leads to too many wiggles,
whereas lower frequencies (periods 300-600\,s) yield too few wiggles.
Note that it is more difficult to determine the periods since these
often exceed the spicule lifetimes
\citep[see also][]{2007Sci...318.1574D}, 

In summary, our Monte-Carlo analysis provides reasonably well-defined
constraints.  In order to reproduce the appearance of our limb spectra
the torsional and swaying motions should be of order 30 and 15
km\,s$^{-1}$, respectively, with periodicities of order $100-300$s.

The next issue is whether the transverse swaying and torsional
modulations represent propagating waves.  Since our limb spectra sample only
one height above the limb at a time, we address this question with 
CRISP imaging spectroscopy in H$\alpha$.  The ultra-high cadence
sequence permits the construction of Doppler images by subtracting the
images taken in the red and blue wings (corresponding to $\pm 55$
km\,s$^{-1}$ Dopplershift).  In these we often observe very fast
propagation of the Doppler signal.  An example is shown in
Figure~\ref{fig:dop} where the apparent phase speed is
285~km\,s$^{-1}$.  Values in the range 100-300 km\,s$^{-1}$ are common.
This is the order of magnitude expected for the Alfv\'en speed in
structures with densities of order $10^{10}$~cm$^{-3}$
\citep{1968SoPh....3..367B} 
and magnetic field strengths of order 10-30~Gauss 
\citep{2010ApJ...708.1579C}.

Determining whether these Dopplershift modulations represent swaying
or torsional motion is not straightforward. In our Dopplergrams, torsional modulation
will show up as a black-and-white pattern across a spicule only when the
sum of the swaying motion and projection of the field-aligned flow
happens to be zero.  The large offsets from line center in Figure~\ref{fig:lx} suggest
that such cancellation does not occur often.  Direct separation of
the torsional and swaying modes is impeded by the combination of very high phase
speed, sparse wavelength sampling, and relatively low cadence, i.e.,
lack of simultaneity between the red- and blue-wing samplings.
Nevertheless, the black-and-white pattern of the spicule at $t \!=\!
34.32$\,s in Figure~\ref{fig:dop} seems direct evidence of
significant torsional motion.  Its propagation at about 300~km\,s$^{-1}$ provides
further support that the observed rotational motions are a signature
of torsional Alfv\'en waves that propagate outward along spicules.

\begin{figure*}[!t]
\includegraphics[width=\columnwidth]{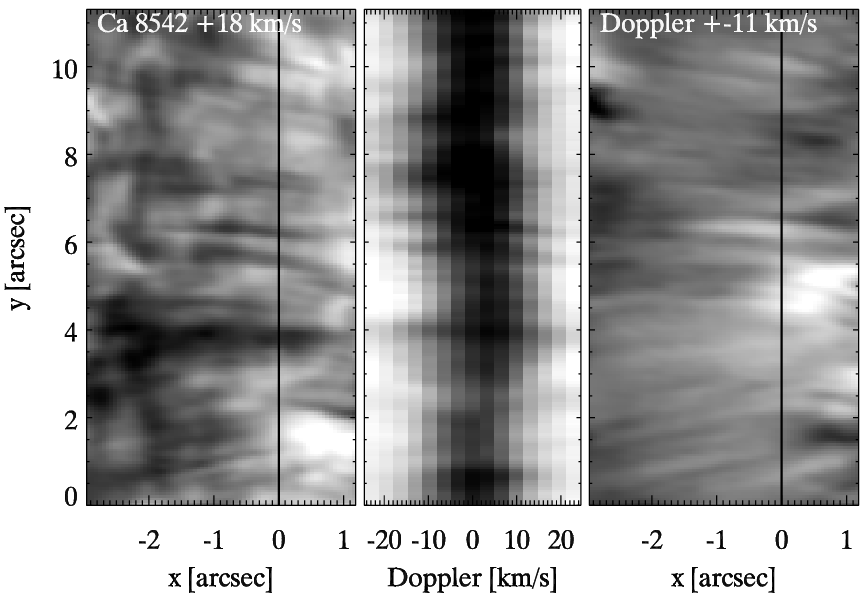}
\includegraphics[width=\columnwidth]{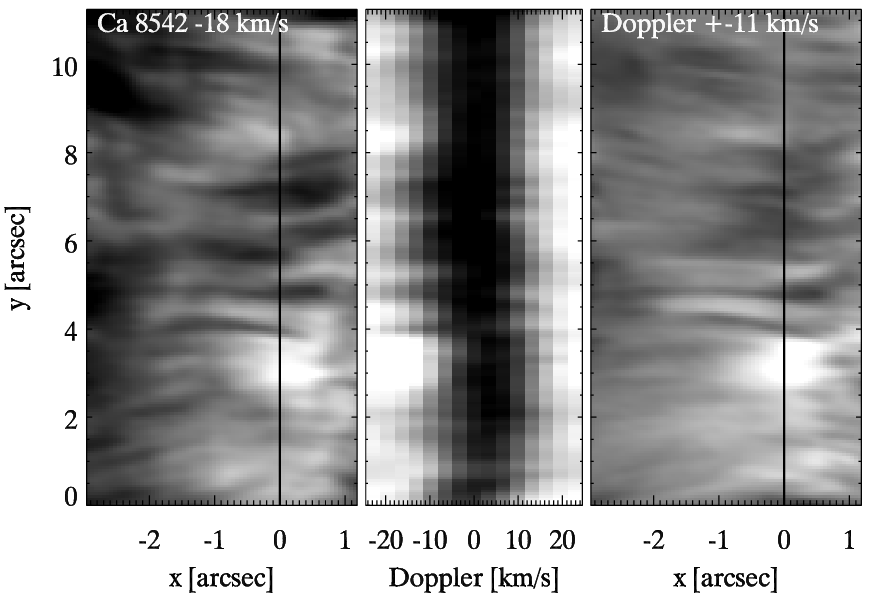}\par
\includegraphics[width=\columnwidth]{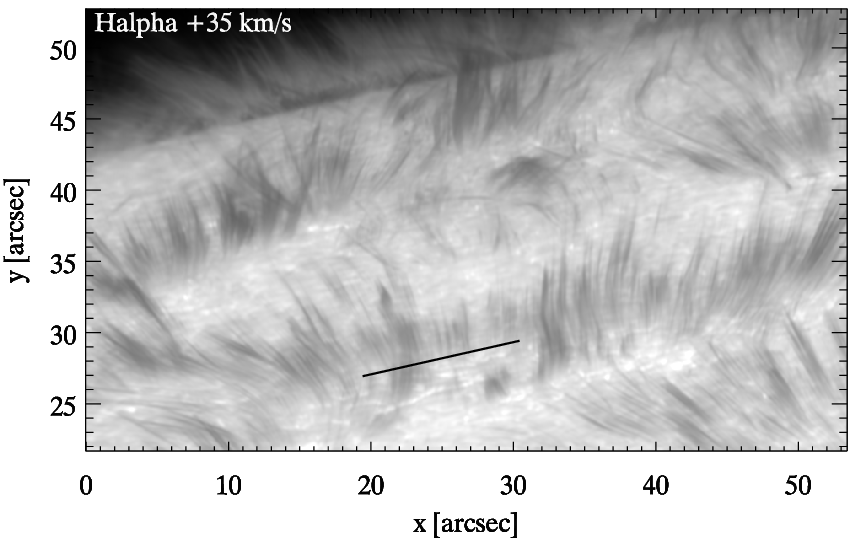}
\includegraphics[width=\columnwidth]{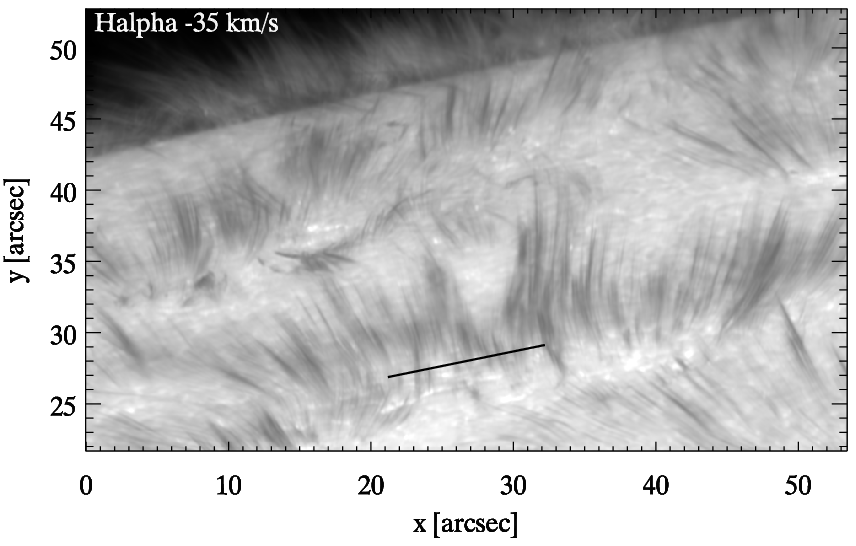}
\caption[]{Two examples of CRISP Ca\,{\sc II}\,854.2\,nm spectral
  behavior close to the limb which is to the left (in top row). The
  slit-like sampling crossed a
  bush of near-limb spicules as shown by the black line in the H$\alpha$
  context images in the lower row.  In the top row, each Ca\,{\sc II}
  $\lambda\!-\!y$ diagram is accompanied by two Ca\,{\sc II} slit-jaw
  images that have the location of the slit marked by a black line, at
  left a wing image, at right a Doppler image. The Ca\,{\sc II}
  spectrograms and images were taken 17\,s apart, the righthand trio
  first. The H$\alpha$ images were taken within a few seconds of each
  other 34\,s later.}
\label{fig:crisp_lambdax}
\end{figure*}

\section{Discussion}
\label{sec:discussion}

Our observations and analysis lead to the conclusion that spicules
undergo three different types of motions at the same time:
field-aligned flow aligned with the spicule axis, swaying that moves
the spicule as a whole transverse to its axis, and torsional motion
around its axis.  The superposition of these motions complicates the
interpretation of spicule spectroscopy and imaging considerably.  Our
Monte Carlo simulations served to disentangle them and indicate that
our observations are compatible with field-aligned flows of order
50-100~km\,s$^{-1}$, swaying of order 15-20~km\,s$^{-1}$, and
torsional motion of order 25-30~km\,s$^{-1}$.  We also found evidence
that the swaying and torsional motions are both signatures of
Alfv\'enic waves with periods of order 100-300\,s with propagation
along the spicule axis at phase speeds as high as
100-300~km\,s$^{-1}$.  Our data are inadequate to decide
whether only upward propagation occurs, or whether propagation in both
directions or even standing waves are as important for the torsional
motions as they seem to be for swaying motions \citep{2011ApJ...736L..24O}.




Our results impact several long-standing issues. First, any
interpretation of observed spicular motion is likely wrong if not all
three kinetic modes are properly accounted for.  The classic reports
of field-aligned outflows of order 20-30 km\,s$^{-1}$ from Doppler
shifts at the limb 
\citep[e.g.,][]{1972ARA&A..10...73B} 
did not include torsional and swaying contributions. 
This introduces additional uncertainty in the
historically reported values for field-aligned spicular flows, which are
already plagued by poor spatio-temporal resolution.



Second, juxtaposition of the three modes of motion explains the
appearance of spicular features in the blue and red wings of Ca\,II\,854.2~nm
and H$\alpha$ close to the limb.  Such images show rows (``bushes'')
of near-vertical absorption features that are comparable to the
on-disk ``rapid blue excursions'' (RBEs) of
\citet{2009ApJ...705..272R}. 
When these are compared between near-simultaneous red and blue
outer-wing samplings, the red-shifted and blue-shifted features appear
with remarkably similar, though not identical, morphology and
orientation.  Precise
fine-scale overlap is rare but the bush patterns often appear roughly
the same.  An H$\alpha$ example pair is shown in the lower panels of
Figure~\ref{fig:crisp_lambdax}.  Such outer-wing similarity has been
interpreted as presence of similar field-aligned flows pointing
towards and away from the observer \citep{1972ARA&A..10...73B}, but the asymmetry
of near-limb line-of-sight projection should then cause large
morphological inequality.
It is much more likely that large-amplitude transverse motions
dominate the visibility close to the limb of outer-wing features and that
they cause morphological equality through near-cospatial and
alternating red and blue Dopplershift modulation.  
The upper panels of Figure~\ref{fig:crisp_lambdax} indeed show
spicular Doppler tilts in the central spectrum panels, much like the
off-limb ones in Figure~\ref{fig:lx}.  This suggests that on-disk RBEs
also undergo torsional motion.

Our observations confirm the speculation of
\citet{2011A&A...532L...9C} 
that torsional motions likely 
occur on smaller scales than their SUMER observations of 
macrospicules and explosive events \citep[or the so-called swirls, see][]{2009A&A...507L...9W}.
Our results also support
the early work by \citet{1970SoPh...13..301A} and suggestions by 
\citet{1968SoPh....5..131P} 
that upward propagating Alfv\'en waves explain
indications of high-speed propagation of Doppler signals along
spicules. The presence and/or observed periods
of propagating torsional Alfv\'en waves can in principle provide information about the thermodynamic
and magnetic properties of the guiding structures
\citep[e.g.,][]{2010ApJ...709.1297R,2010ApJ...714.1637V}. Such idealized models are not yet a good
 representation of actual, highly dynamic spicules but may
 illustrate the potential diagnostic value of spicular wave observations.

The ubiquity of torsional spicule motion also provides support for
scenarios of spicule formation in which non-linear coupling of
torsional Alfv\'en waves to other wave modes on expanding flux tubes leads to significant
field-aligned flows that drive the plasma upwards
\citep{1982SoPh...75...35H, 1984ApJ...285..843S,
  1998A&A...338..729D, 
  1999ApJ...514..493K,2010ApJ...710.1857M}.  However, none of these
models can explain both the large observed
field-aligned flows and the rapid heating to at least
transition-region temperatures. This suggests that while torsional
motions may play a role in providing upward momentum in spicules,
other components of spicule formation are still missing. 

While our Monte Carlo simulation assumes the presence of only the
$m\!=\!0$ torsional mode, the observations show also evidence for
more complex torsional modes with $m \!>\! 0$ (see example at
y=34\arcsec\ in top left panel of Figure~\ref{fig:lx}). 
Our results thus may provide support for the recent hypothesis \citep{2011ApJ...736....3V,2012ApJ...746...81A}
that complex torsional motions are generated in the interaction between
convective motions and photospheric flux tubes \citep[see, e.g., the
ubiqitous vorticy reported in recent
simulations by][]{2011A&A...533A.126M}, and that they may be
responsible for
heating of the chromosphere and corona.

The observed torsional motions imply that spicules, given their
ubiquity, play a major role in the transport of helicity through the
solar atmosphere. In addition, our results suggest that the energy
flux carried by Alfv\'enic motions into the corona and heliosphere may
be twice as large as the previous estimates that were based on swaying
motions only 
\citep{2007Sci...318.1574D, 
2011Natur.475..477M}. 


  %


\acknowledgments
The Swedish 1-m Solar Telescope is operated by the Institute for Solar
Physics of the Royal Swedish Academy of Sciences in the Spanish
Observatorio del Roque de los Muchachos of the Instituto de
Astrof\'{\i}sica de Canarias. 
D.H.~Sekse, T.~Leifsen, and G.~Vissers acquired the 11-jun-2011 observations. 
This research was supported by the Research Council of Norway and
by the European Union through the ERC Advanced Grant "Physics of the Solar Chromosphere".
B.D.P. was supported through NASA grants NNX08BA99G, NNX08AH45G and
NNX11AN98G.  
  %
The
authors are grateful to Alan Title, Scott McIntosh and Neil Murphy for
insightful discussions.


\end{document}